\documentclass{LMCS}

\usepackage{graphicx}
\usepackage{color}
\usepackage{latexsym}
\usepackage{hyperref}

\newcommand{\impliesarrow}{\Rightarrow}
\newcommand{\True}{\mbox{\sc True}}
\newcommand{\False}{\mbox{\sc False}}

\newcommand{\Sp}[2]{\mbox{sp}_{#1}(#2)}
\newcommand{\Scp}[2]{\mbox{scp}_{#1}(#2)}
\newcommand{\Hi}[2]{\mbox{hi}_{#1}(#2)}
\newcommand{\Wp}[2]{\mbox{wp}_{#1}(#2)}

\newcommand{\AbsProg}[2]{\Pi'}
\newcommand{\Spop}{\mbox{sp}}
\newcommand{\Itp}[1]{\hat{#1}}
\newcommand{\Approx}[1]{\hat{#1}}
\newcommand{\Abs}[1]{\bar{#1}}
\newcommand{\ItpFun}{\mbox{\sc Itp}}
\newcommand{\Next}[1]{\dot{#1}}
\newcommand{\Prime}[2]{#1^{\langle #2 \rangle}}
\newcommand{\Tof}[1]{T(#1)}
\newcommand{\Blast}{{\sc Blast}}
\newcommand{\Slam}{{\sc Slam}}
\newcommand{\Antepos}[1]{{#1}^+}
\newcommand{\Anteneg}[1]{{#1}^-}
\newcommand{\Pivot}[1]{\mbox{\sc Pivot}(#1)}
\newcommand{\asterisk}{*}

\def\doi{3 (4:1) 2007}
\lmcsheading%
{\doi}
{1--17}
{}
{}
{Jun.~12, 2006}
{Nov.~\phantom{0}1, 2007}
{}

\begin{document}

\title{Interpolant-based Transition Relation Approximation\rsuper *}

\author[R.~Jhala]{Ranjit Jhala\rsuper a}	
\address{{\lsuper a}University of California, San Diego \\
9500 Gilman Drive \\
La Jolla, CA 92093}	
\email{jhala@cs.ucsd.edu}  

\author[K.~L.~McMillan]{Kenneth L. McMillan\rsuper b}	
\address{{\lsuper b}Cadence Berkeley Laboratories \\
1995 University Ave., Suite 460\\
Berkeley, CA  94704
}	
\email{mcmillan@cadence.com}


\titlecomment{{\lsuper*}A preliminary version of this paper appeared in the
proceedings of CAV 2005.}

\begin{abstract}
  In predicate abstraction, exact image computation is problematic,
  requiring in the worst case an exponential number of calls to a
  decision procedure. For this reason, software model checkers
  typically use a weak approximation of the image.  This can result in
  a failure to prove a property, even given an adequate set of
  predicates. We present an interpolant-based method for strengthening
  the abstract transition relation in case of such failures. This
  approach guarantees convergence given an adequate set of predicates,
  without requiring an exact image computation. We show empirically
  that the method converges more rapidly than an earlier method based
  on counterexample analysis.
\end{abstract}
\maketitle

\section{Introduction}
Predicate abstraction~\cite{GS97} is a technique commonly used in
software model checking in which an infinite-state system is
represented abstractly by a finite-state system whose states are the
truth valuations of a chosen set of predicates. The
reachable state set of the abstract system corresponds to the
strongest inductive invariant of the infinite-state system expressible
as a Boolean combination of the given predicates.

The primary computational difficulty of predicate abstraction is the
\emph{abstract image} computation. That is, given a set of predicate
states (perhaps represented symbolically) we wish to compute the set of
predicate states reachable from this set in one step of the abstract system. This can be done
by enumerating the predicate states, using a suitable decision procedure to
determine whether each state is reachable in one step. However, since
the number of decision procedure calls is exponential in the number of
predicates, this approach is practical only for small predicates
sets. For this reason, software model checkers, such as \Slam~\cite{SLAM}
and \Blast~\cite{BLAST1} typically use weak approximations of the
abstract image. For example, the Cartesian image approximation is the
strongest cube over the predicates that is implied at the next time.
This approximation loses all information about predicates that are
neither deterministically true nor deterministically false at the next
time. Perhaps surprisingly, some properties of large programs, such as
operating system device drivers, can be verified with this weak
approximation~\cite{SLAM,POPL04}. Unfortunately, as we will observe, this
approach fails to verify properties of even very simple programs, if
the properties relate to data stored in arrays.

This paper introduces an approach to approximating the transition
relation of a system using Craig interpolants derived from proofs
of bounded model checking instances. These interpolants are formulas
that capture the information about the transition
relation of the system that was deduced in proving the property in a bounded sense.
Thus, the transition relation
approximation we obtain is tailored to the property we are trying to
prove. Moreover, it is a
formula over only state-holding variables.  Hence, for abstract
models produced by predicate abstraction, the
approximate transition relation is a purely propositional formula,
even though the original transition relation is characterized by a
first-order formula.  Thus, we can apply well-developed Boolean image
computation methods to the approximate system, eliminating the
need for a decision procedure in the image computation.
By iteratively refining the approximate transition relation we can
guarantee convergence, in the sense that whenever the chosen
predicates are adequate to prove the property, the approximate
transition relation is eventually strong enough to prove the property.

The reader should bear in mind that there are two kinds of abstraction
occurring here. The first is \emph{predicate abstraction}, which
produces an abstract transition system whose state-holding variables are
propositional.  The second is \emph{transition relation
  approximation}, which weakens the abstract transition formula,
yielding a purely propositional approximate transition formula.  To
avoid confusion, we will always refer to the former as
\emph{abstraction}, and the latter as \emph{approximation}. The
techniques presented here produce an exact reachability result \emph{for the
abstract model}. However, we may still fail to prove unreachability if
an inadequate set of predicates is chosen for the abstraction.

After beginning with an overview of closely related work (Section~\ref{sec:related}), 
we introduce some notation and definitions related to modelling
infinite-state systems symbolically and briefly describe the method 
of deriving interpolants from proofs (Section~\ref{sec:prelim}). 
Next, we introduce the basic method of transition relation approximation 
using interpolants (Section~\ref{sec:transinterp}). 
In the following section, we discuss a number of optimizations of this basic method that are
particular to software verification, including a new method to strengthen
interpolants that makes convergence more rapid
(Section~\ref{sec:strengthen}). Finally, we present an experimental 
presents an experimental comparison of the interpolation method with
the Das and Dill method (Section~\ref{sec:results}).

\section{Related work}\label{sec:related}
The chief alternative to iterative approximation is to produce an
exact propositional characterization of the abstract transition
relation.  For example the method of~\cite{SBC2003} uses small-domain
techniques to translate a first-order transition formula into a
propositional one that is equisatisfiable over the state-holding
predicates. However, this translation introduces a large number of
auxiliary Boolean variables, making it impractical to use BDD-based
methods for image computation. Though SAT-base Boolean quantifier elimination
methods can be used, the effect is still essentially to enumerate the
states in the image. By contrast, the interpolation-based method
produces an approximate transition relation with no auxiliary Boolean
variables, allowing efficient use of BDD-based methods.

The most closely related method is that of Das and Dill~\cite{DBLP:conf/lics/DasD01}.
This method analyzes abstract counterexamples (sequences of predicate
states), refining the transition relation approximation in such a way
as to rule out infeasible transitions. This method is effective, but
has the disadvantage that it uses a specific counterexample and does not
consider the property being verified. Thus it can easily generate
refinements not relevant to the property. The
interpolation-based method does not use abstract
counterexamples. Rather, it generates facts relevant to proving the
given property in a bounded sense.  Thus, it tends to generate more
relevant refinements, and as a result converges more rapidly.

In~\cite{POPL04}, interpolants are used to choose new predicates
to refine a predicate abstraction. Here, we use interpolants to
refine an approximation of the abstract transition relation for
a given set of predicates.

\section{Preliminaries}\label{sec:prelim}

Let~$S$ be a first-order signature, consisting of
individual variables and uninterpreted $n$-ary functional and propositional
constants. A \emph{state formula} is a first-order formula over~$S$,
(which may include various interpreted symbols, such as~$=$ and~$+$).
We can think of a state formula $\phi$ as representing a set of states, namely,
the set of first-order models of~$\phi$. We will express the proposition
that an interpretation~$\sigma$ over $S$ models $\phi$ by $\phi[\sigma]$.

We also assume a first-order signature~$S'$, disjoint from~$S$,
and containing for every symbol $s\in S$, a unique symbol~$s'$ of the same type.
For any
formula or term $\phi$ over~$S$, we will use $\phi'$ to represent the result
of replacing every occurrence of a symbol $s$ in $\phi$ with $s'$. Similarly, for
any interpretation $\sigma$ over~$S$, we will denote by $\sigma'$ the interpretation
over~$S'$ such that $\sigma' s' = \sigma s$.
A \emph{transition formula} is a first-order formula over
$S \cup S'$. We think of a transition formula~$T$ as representing a set of
state pairs, namely the set of pairs $(\sigma_1,\sigma_2)$, such that
$\sigma_1 \cup \sigma'_2$ models $T$. Will will express the proposition
that $\sigma_1 \cup \sigma'_2$ models $T$ by $T[\sigma_1,\sigma_2]$.

The \emph{strongest postcondition} of a state formula~$\phi$ with
respect to transition formula~$T$, denoted $\Sp{T}{\phi}$, is the
strongest proposition~$\psi$ such that $\phi \wedge T$ implies~$\psi'$. 
We will also refer to this as the \emph{image} of $\phi$
with respect to $T$. Similarly, the \emph{weakest precondition} of a state
formula~$\phi$ with respect to transition formula~$T$, denoted
$\Wp{T}{\phi}$ is the weakest proposition~$\psi$ such that $\psi
\wedge T$ implies~$\phi'$. 

A \emph{transition system} is a pair $(I,T)$, where $I$ is a state
formula and $T$ is a transition formula. 
Given a state formula $\psi$, we will say that $\psi$ is
\emph{$k$-reachable} in~$(I,T)$ when there exists a sequence of
states $\sigma_0,\ldots,\sigma_k$, such that $I[\sigma_0]$ and for
all $0 \leq i < k$, $T[\sigma_i,\sigma_{i+1}]$, and $\psi[\sigma_k]$.
Further,~$\psi$ is \emph{reachable} in~$(I,T)$ if it is $k$-reachable for some~$k$.
We will say that~$\phi$ is an \emph{invariant} of $(I,T)$ when $\neg\phi$ is
not reachable in $(I,T)$.  A state formula~$\phi$ is an
\emph{inductive invariant} of $(I,T)$ when $I$ implies $\phi$ and
$\Sp{T}{\phi}$ implies $\phi$ (note that an inductive invariant is
trivially an invariant).

\subsection{Bounded model checking}

The fact that $\psi$ is $k$-reachable in~$(I,T)$
can be expressed symbolically. For any symbol $s$, and natural number~$i$, we will use the notation
$\Prime{s}{i}$ to represent the symbol~$s$ with $i$ primes added. Thus, $\Prime{s}{3}$ is $s'''$.
A symbol with~$i$ primes will be used to represent the value of that symbol at time~$i$.
We also extend this notation to formulas. Thus, the formula~$\Prime{\phi}{i}$ is
the result of adding~$i$ primes to every uninterpreted symbol in~$\phi$.

Now, assuming $T$ is total, the state formula $\psi$ is $k$-reachable in~$(I,T)$ exactly when this
formula is consistent:
\[\Prime{I}{0} \wedge \Prime{T}{0} \wedge \cdots \Prime{T}{k-1} \wedge \Prime{\psi}{k} \]
We will refer to this as a \emph{bounded model checking} formula~\cite{biere99symbolic},
since by testing satisfiability of such formulas, we can determine the
reachability of a given condition within a bounded number of steps.

\subsection{Interpolants from proofs}

Given a pair of formulas $(A,B)$, such that $A\wedge B$ is inconsistent,
an \emph{interpolant} for $(A,B)$ is a formula $\Itp{A}$ with the
following properties:
\begin{itemize}
\item $A$ implies $\Itp{A}$,
\item $\Itp{A} \wedge B$ is unsatisfiable, and
\item $\Itp{A}$ refers only to the common symbols of $A$ and $B$.
\end{itemize}
Here, ``symbols'' excludes symbols such as $\wedge$ and $=$ that are
part of the logic itself. Craig showed that for first-order formulas,
an interpolant always exists for inconsistent formulas~\cite{Craig}.  Of more
practical interest is that, for certain proof systems, an interpolant
can be derived from a refutation of $A\wedge B$ in linear time. For
example, a purely propositional refutation of $A\wedge B$ using the
resolution rule can be translated to an interpolant in the form of a Boolean circuit having the
same structure as the proof~\cite{Kraj,Pudlak}.

In~\cite{TCS} it is shown that linear-size interpolants
can be derived from refutations in a first-order theory with
uninterpreted function symbols and linear arithmetic. This
translation has the property that whenever $A$ and $B$ are quantifier-free,
the derived interpolant~$\Itp{A}$ is also quantifier-free.\footnote{
Note that the Craig theorem does not guarantee the existence
of quantifier-free interpolants. In general this depends on the
choice of interpreted symbols in the logic.} We will exploit this
property in the sequel.

Heuristically, the chief advantage of interpolants derived from
refutations is that they capture the facts that the prover derived
about~$A$ in showing that $A$ is inconsistent with~$B$. Thus, if the
prover tends to ignore irrelevant facts and focus on relevant ones,
we can think of interpolation as a way of filtering out irrelevant
information from~$A$. 

For the purposes of this paper, we must extend the notion of interpolant
slightly.
That is, given an indexed
set of formulas $A = \{a_1,\ldots,a_n\}$ such that $\bigwedge A$ is inconsistent,
a \emph{symmetric
interpolant} for $A$ is an indexed set of formulas $\Itp{A} = \{\Itp{a}_1,\ldots,\Itp{a}_n\}$
such that each $a_i$ implies $\Itp{a}_i$, and $\bigwedge\Itp{A}$ is inconsistent, and
each $\Itp{a}_i$ is over the symbols common to $a_i$ and $A\setminus a_i$.
We can construct a symmetric interpolant for $A$ from a refutation of
$\bigwedge A$ by simply letting $\Itp{a}_i$ be the interpolant derived from
the given refutation for the pair $(a_i,\bigwedge A\setminus a_i)$.  As long as
all the individual interpolants are derived \emph{from the same proof}, we
are guaranteed that their conjunction is inconsistent. In the sequel, if~$\Itp{A}$
is a symmetric interpolant for~$A$, and the elements of~$A$ are not explicitly indexed,
we will use the notation $\Itp{A}(a_i)$ to refer to $\Itp{a_i}$.

\section{Transition relation approximation}\label{sec:transinterp}
Because of the expense of image computation in symbolic
model checking, it is often beneficial to abstract the transition
relation before model checking, removing information that
is not relevant to the property to be proved. Some examples of
techniques for this purpose are~\cite{clarke00counterexampleguided,absref}.

In this paper, we introduce a method of approximating the transition
relation using bounded model checking and symmetric interpolation.
Given a transition system $(I,T)$ and a state formula~$\psi$ that we wish to
prove unreachable, we will use interpolation to refine an approximation
$\Approx{T}$ of the transition relation~$T$, such that $T$ implies $\Approx{T}$.
The initial approximation is just $\Approx{T} = \True$.

We begin the refinement loop by attempting to verify the unreachabilty
of~$\psi$ in the approximate system $(I,\Approx{T})$, using an
appropriate model checking algorithm.  If~$\psi$ is found to be
unreachable in $(I,\Approx{T})$, we know it is unreachable in the
stronger system~$(I,T)$. Suppose, on the other hand that~$\psi$ is found to be $k$-reachable in $(I,\Approx{T})$.
It may be that in fact~$\psi$ is $k$-reachable in~$(I,T)$, or
it may be that $\Approx{T}$ is simply too weak an approximation to refute this.
To find out, we will use bounded model checking.

That is, we construct the following set of formulas:
\[A \doteq \{\Prime{I}{0},\Prime{T}{0},\ldots,\Prime{T}{k-1},\Prime{\psi}{k}\} \]
Note that $\bigwedge A$ is exactly the bounded model checking formula
that characterizes $k$-reachability of~$\psi$ in $(I,T)$.
We use a decision procedure to determine satisfiability of~$\bigwedge A$.
If it is satisfiable, $\psi$ is reachable and we are done. If not, we obtain
from the decision procedure a refutation of~$\bigwedge A$. From this,
we extract a symmetric interpolant~$\Itp{A}$.
Notice that for each $i$ in $0\ldots{k-1}$, $\Itp{A}(\Prime{T}{i})$ is a formula
implied by $\Prime{T}{i}$, the transition formula shifted to time~$i$. Let us shift these formulas
back to time~$0$, thus converting them to transition formulas. That is, for $i=0\ldots k-1$, let:
\[\Itp{T}_i \doteq \Prime{(\Itp{A}(\Prime{T}{i}))}{-i}\]
where we use~$\Prime{\phi}{-i}$ to denote removal of~$i$ primes from~$\phi$, when feasible.
We will call these formulas the \emph{transition interpolants}.
From the properties of symmetric interpolants, we know the bounded model
checking formula
\[I_0 \wedge \Prime{\Itp{T}_0}{0} \wedge \cdots \Prime{\Itp{T}_{k-1}}{k-1} \wedge \psi_k \]
is unsatisfiable. Thus we know that the 
conjunction of the transition interpolants~$\bigwedge_i \Itp{T}_i$ 
admits no path of~$k$ steps from~$I$ to~$\psi$. We now compute a refined approximation
$\Next{T} \doteq \Approx{T} \wedge \bigwedge_i \Itp{T}_i$.
This becomes our approximation~$\Approx{T}$ in the next  iteration
of the loop.
This procedure is summarized
in Figure~\ref{fig:loop}.
\begin{figure}[tb]
  \centering
  \begin{tabbing}
    mmmm\=mmmm\=mmmm\=mmmm\=mmmm\=mmmm\=\kill
    $\Approx{T} \leftarrow \True$\\
    \textbf{repeat}\\
    \> if $\psi$ unreachable in $(I,\Approx{T})$, return ``unreachable''\\
    \> else, if $\psi$ reachable in $k$ steps in $(I,\Approx{T})$\\
    \> \> $A \leftarrow \{\Prime{I}{0},\Prime{T}{0},\ldots,\Prime{T}{k-1},\Prime{\psi}{k}\}$\\
    \> \> if $\bigwedge A$ satisfiable, return ``reachable in $k$ steps''\\
    \> \> else\\
    \> \> \> $\Itp{A} \leftarrow \ItpFun(A)$\\
    \> \> \> $\Approx{T} \leftarrow \Approx{T} \wedge \bigwedge_{i=0}^{k-1} \Prime{(\Itp{A}(\Prime{T}{i}))}{-i}$\\
    \textbf{end repeat}
  \end{tabbing}
  \caption{Interpolation-based transition approximation loop. Here, $\ItpFun$ is a
    function that computes a symmetric interpolant for a set of formulas.}
  \label{fig:loop}
\end{figure}
Notice that at each iteration, the refined approximation~$\Next{T}$ is strictly
stronger than~$\Approx{T}$, since $\Approx{T}$ allows a counterexample of~$k$ steps, but
$\Next{T}$ does not. Thus, for finite-state systems, the loop must terminate.
This is simply because we cannot strengthen a formula with a finite number
of models infinitely.

The approximate transition formula~$\Itp{T}$ has two principle
advantages over~$T$.  First, it contains only facts about the
transition relation that were derived by the prover in resolving the
bounded model checking problem. Thus it is in some sense an
abstraction of~$T$ relative to~$\psi$. Second,~$\Itp{T}$ contains only
state-holding symbols. We will say that a symbol~$s\in S$ is
\emph{state-holding} in $(I,T)$ when~$s$ occurs in~$I$, or~$s'$ occurs
in $T$. In the bounded model checking formula, the only symbols in
common between $\Prime{T}{i}$ and the remainder of the formula are of the
form~$\Prime{s}{i}$ or $\Prime{s}{i+1}$, where~$s$ is state-holding. Thus, the
transition interpolants~$\Itp{T}_i$ contain only state-holding symbols
and their primed versions.

The elimination of the non-state-holding symbols by interpolation has
two potential benefits.  First, in hardware verification there are
usually many non-state-holding symbols representing inputs of the
system. These symbols contribute substantially to the cost of the
image computation in symbolic model checking. Second, for this
paper, the chief benefit is in the case when the state-holding symbols are
all propositional (\emph{i.e.}, they are propositional constants).  In
this case, even if the transition relation~$T$ is a first-order
formula, the approximation~$\Approx{T}$ is a propositional formula.
The individual variables and function symbols are eliminated by
interpolation.  Thus we can apply well-developed Boolean methods for
symbolic model checking to the approximate system.  In the next
section, we will apply this approach to predicate abstraction.

\section{Application to predicate abstraction}

Predicate abstraction~\cite{GS97} is a technique commonly used in
software model checking in which the state of an infinite-state system
is represented abstractly by the truth values of a chosen set of
predicates~$P$. The method computes the strongest inductive
invariant of the system expressible as a Boolean combination of these
predicates.

Let us fix a concrete transition system $(I,T)$
and a finite set of state formulas~$P$ that we will refer to simply as ``the
predicates''.  We assume a finite set $V \subset S$ of uninterpreted propositional
symbols not occurring in~$I$ or~$T$. The set~$V$ consists of a symbol $v_p$ for
every predicate $p\in P$. We will
construct an abstract transition system~$(\Abs{I},\Abs{T})$ whose states
are the minterms over~$V$. 
To relate the abstract and concrete systems, we define a concretization
function~$\gamma$. Given a formula over $V$, $\gamma$ replaces every occurrence
of a symbol $v_p$ with the corresponding predicate $p$. Thus, if $\phi$ is a Boolean combination over~$V$,
$\gamma(\phi)$ is the same combination of the corresponding predicates in~$P$.

For the sake of simplicity, we assume that the initial condition $I$ is a Boolean
combination of the predicates. Thus we choose~$\Abs{I}$ so that $\gamma(\Abs{I}) = I$.
We define the abstract transition relation~$\Abs{T}$ such that,
for any two minterms $s,t \in 2^V$, we have $\Abs{T}[s,t]$ exactly
when $\gamma(s) \wedge T \wedge \gamma(t)'$ is consistent. In other
words, there is a transition from abstract state~$s$ to abstract state
$t$ exactly when there is a transition from a concrete state
satisfying~$\gamma(s)$ to a concrete state satisfying~$\gamma(t)$.

We can easily show by induction on the number of steps that if a
formula~$\psi$ over $V$ is unreachable in~$(\Abs{I},\Abs{T})$ then
$\gamma(\psi)$ is unreachable in $(I,T)$ (though the converse does not
hold). 
To allow us to check whether a given~$\psi$ is in fact reachable in the abstract system,
we can express the abstract transition relation symbolically~\cite{SBC2003}. The abstract transition
relation can be expressed as
\[\Abs{T} \doteq \left(\left({\textstyle \bigwedge_{p\in P}} (v_p \iff p)\right) \wedge T \wedge \left({\textstyle \bigwedge_{p\in P}} (p' \iff v'_p)\right)
  \right) \downarrow (V\cup V')\]
where $Q\downarrow W$ denotes the ``hiding'' of non-$W$ symbols in $Q$ by renaming them to fresh
symbols in $S$. Hiding the concrete symbols in this way takes the place of existential quantification.
Notice that, under this definition, the state-holding symbols of $(\Abs{I},\Abs{T})$
are exactly~$V$. Moreover, for any two minterms $s,t \in 2^V$, the formula $s\wedge \Abs{T} \wedge t'$
is consistent exactly when $\gamma(s) \wedge T \wedge \gamma(t)'$ is consistent. Thus, $\Abs{T}$ characterizes
exactly the transitions of our abstract system.

To determine whether~$\psi$ is reachable in this system using the
standard ``symbolic'' approach, we would compute the reachable states~$R$
of the system as the limit of the following recurrence:
\begin{eqnarray*}
  R_0 &\doteq& \Abs{I}\\
  R_{i+1} &\doteq& R_i \vee \Sp{\Abs{T}}{R_i}
\end{eqnarray*}
The difficulty here is to compute the image~$\Spop_{\Abs{T}}$. We cannot apply standard
propositional methods for image computation, since the transition
formula~$\Abs{T}$ is not propositional. We can compute $\Sp{\Abs{T}}{\phi}$ as
the disjunction of all the minterms $s\in2^V$ such that $\phi \wedge \Abs{T} \wedge s'$
is consistent. However, this is quite expensive in practice, since it requires an
exponential number of calls to a theorem prover. In~\cite{SBC2003}, this is avoided
by translating~$\Abs{T}$ into a propositional formula that is 
equisatisfiable with~$\Abs{T}$ over~$V\cup V'$. This makes it possible to use well developed Boolean image computation
methods to compute the abstract strongest postcondition. 
Nonetheless, because the translation introduces a large number
of free propositional variables, the standard approaches to image computation
using Binary Decision Diagrams (BDD's) were found to be inefficient.
Alternative methods based on enumerating the satisfiable assignments
using a SAT solver were found to be more effective, at least for
small numbers of predicates. However, this method is still essentially
enumerative. Its primary advantage is that information learned by
the solver during the generation of one satisfying assignment can
be reused in the next iteration.

Here, rather than attempting to compute images exactly in the abstract
system, we will simply observe that state-holding symbols of the
abstraction $(\Abs{I},\Abs{T})$ are all propositional. Thus, the
interpolation-based transition relation approximation method of the
previous section reduces the transition relation to a purely
propositional formula. Moreover, it does this without introducing
extraneous Boolean variables. Thus, we can apply standard BDD-based
model checking methods to the approximated system
$(I,\Approx{T})$ without concern that non-state-holding Boolean
variables will cause a combinatorial explosion. Finally, termination
of the approximation loop is guaranteed because the abstract state space is finite.

\section{Software model checking}

In model checking sequential deterministic programs, we can make some
significant optimizations in the above method.

\subsection{Path-based approximation}

The first optimization is to treat the program counter explicitly, rather than
modeling it as a symbolic variable. The main advantage of this is that it will allow
us to apply bounded model checking only to particular program paths (\emph{i.e.},
sequences of program locations) rather than to the program as a whole. 

We will say that a \emph{program} $\Pi$ is a pair $(L,R)$, where $L$
is a finite set of \emph{locations}, and $R$ is a finite set of
\emph{operations}. An operation is a triple $(l,T,l')$ where $T$
is a transition formula, $l \in L$ is the entry location of the statement, and
$l' \in L$ is the exit location of the statement.

A \emph{path} of program $\Pi$ from location $l_0\in L$ to location $l_k\in L$ is a
sequence $\pi \in R^{k-1}$, of the form
$(l_0,T_0,l_1)(l_1,T_1,l_2)\cdots(l_{k-1},T_{k-1},l_k)$. We
say that the path is~\emph{feasible} when there exists a sequence
of states $\sigma_0\cdots\sigma_k$
such that, for all $0 \leq i < k$, we have $T_i[\sigma_i,\sigma_{i+1}]$.
The reachability
problem is to
determine whether program~$\Pi$ has a feasible path from a given  initial
location~$l_0$ to a given final location~$l_f$.

As in the previous section, we assume a fixed set of predicates~$P$,
and a corresponding set of uninterpreted propositional symbols~$V$. 
Using these, we construct an abstract program $\Abs{\Pi} = (L,\Abs{R})$.
For any operation~$r = (l,T,l')$, let the abstract operation~$\Abs{r}$ be
$(l,\Abs{T},l')$, where, as before
\[\Abs{T} \doteq \left(\left({\textstyle \bigwedge_{p\in P}} (v_p \iff p)\right) \wedge T \wedge \left({\textstyle \bigwedge_{p\in P}} (p' \iff v'_p)\right)\right) \downarrow (V\cup V')\]
The abstract operation set~$\Abs{R}$ is then $\{\Abs{r}\ |\ r\in R\}$. 
We can easily show that if a path $r_0\cdots r_{k-1}$ is feasible,
then the corresponding abstract path~$\Abs{r}_0\cdots \Abs{r}_{k-1}$
is also feasible. Thus if a given location $l_f$ is unreachable from~$l_0$
in the abstract program, it is unreachable from~$l_0$ in the concrete program.

Now we can apply the interpolation-based approximation approach to programs.
We will build an approximate program $\Approx{\Pi} = (L,\Approx{R})$, where
$\Approx{R}$ consists of an operation~$\Approx{r} = (l,\Approx{T},l')$ for
every $\Abs{r} = (l,\Abs{T},l')$ in $\Abs{R}$, such that $\Abs{T}$ implies $\Approx{T}$,
and $\Approx{T}$ is over $V\cup V'$.
Initially, every $\Approx{T}$ is just $\True$.

At every step of the iteration, we use standard model checking methods
to determine whether the approximation $\Approx{\Pi}$ has a feasible path from $l_0$ to
$l_f$. We can do this because the transition formulas~$\Approx{T}$ are all propositional.
If there is no such path, then $l_f$ is not reachable in the concrete program
and we are done. 
Suppose on the other hand that there is such a path $\Approx{\pi} = \Approx{\pi}_0\cdots \Approx{\pi}_{k-1}$.
Let $\Abs{\pi} = \Abs{\pi}_0\cdots \Abs{\pi}_{k-1}$ be the corresponding path of $\Abs{\Pi}$.
We can construct a bounded model checking formula to determine
the feasibility of this path. Using the notation~$\Tof{r}$ to denote the $T$ component
of an operation~$r$, let 
\[A \doteq \{\Prime{\Tof{\Abs{\pi}_i}}{i}\ |\ i \in 0\ldots k-1\}\]
The conjunction~$\bigwedge A$ is consistent exactly when the abstract path~$\Abs{\pi}$
is feasible. Thus, if $\bigwedge A$ is consistent, the abstraction does not prove unreachability
of~$l_f$ and we are done. If it is inconsistent, we construct a symmetric interpolant
$\Itp{A}$ for~$A$. We extract transition interpolants as follows:
\[\Itp{T}_i \doteq \Prime{(\Itp{A}(\Prime{\Tof{\Abs{\pi}_i}}{i}))}{-i}\]
Each of these is implied by the $\Tof{\Abs{\pi}_i}$, the transition formula of the
corresponding abstract operation. We now strengthen our approximate program~$\Approx{\Pi}$
using these transition interpolants. That is, for each abstract operation~$\Abs{r}\in\Abs{R}$,
the refined approximation is~$\Next{r} = (l,\Tof{\Next{r}},l')$
where
\[\Tof{\Next{r}} \doteq \Tof{\Approx{r}} \wedge \left({\textstyle \bigwedge}\{\Itp{T}_i\ |\ {\Abs{\pi}_i = \Abs{r}},\  i\in 0\ldots k-1\}\right) \]
In other words, we constrain each approximate operation $\Approx{r}$ by the set of transition interpolants
for the occurrences of~$\Abs{r}$ in the abstract path~$\Abs{\pi}$. The refined approximate program is
thus $(L,\Next{R})$, where $\Next{R} = \{\Next{r}\ |\ \Abs{r}\in \Abs{R}\}$. 
From the interpolant properties, we can easily show that the refined approximate program
does not admit a feasible path corresponding to~$\Abs{\pi}$.

We continue in this manner until either the model checker determines
that the approximate program~$\Approx{\Pi}$ has no feasible path from $l_0$ to $l_f$, or until bounded
model checking determines that the abstract program~$\Abs{\Pi}$ does have such a feasible path.
This
process must terminate, since at each step $\Approx{\Pi}$ is strengthened, and we
cannot strengthen a finite set of propositional formulas infinitely.

The advantage of this approach, relative to that of section~\ref{sec:transinterp},
is that the bounded model checking formula $\bigwedge A$ only
relates to a single program path. In practice, the refutation of a
single path using a decision procedure is considerably less costly
than the refutation of all possible paths of a given length.
 
As an example of using interpolation to compute an approximate program, Figure~\ref{fig:focisymm} shows a small program with
one path, which happens to be infeasible. The method of~\cite{POPL04} chooses the predicates $x=z$, $a[z]=y$ and $a[z] = y-1$ to
represent the abstract state space. 
Next to each operation in the path is shown the transition interpolant~$\Itp{T}_i$ that was obtained for that operation.
Note that each transition interpolant is
implied by the semantics of the corresponding statement, and that collectively
the transition interpolants rule out the program path (the reader might wish to verify this).
Moreover, the transition interpolant for the first statement, $a[x] \leftarrow y$, is $x=z \impliesarrow a[z]=y$.
This is a disjunction and therefore cannot be inferred by predicate image
techniques that use the Cartesian or Boolean programs approximations. In fact,
the {\sc Blast} model checker cannot rule out this program path. However,
using transition interpolants, we obtain a transition relation approximation
that proves the program has no feasible path from beginning to end.
\begin{figure}[t]
  \centering
  \begin{tabular}[c]{l|l}
    statement & transition interpolant\\
    \hline
    $a[x] \leftarrow y$ & $(x=z)' \impliesarrow (a[z]=y)'$\\
    $y \leftarrow y + 1$ & $(a[z] = y \impliesarrow (a[z] = y-1)') \wedge ((x = z)' \impliesarrow x= z)$\\
    assume $z = x$ & $(a[z] = y-1 \impliesarrow (a[z] = y-1)') \wedge x = z$ \\
    assume $a[z] \neq y-1\ \ \ $ & $a[z] \neq y-1$
  \end{tabular}
  \caption{An infeasible program path, with transition interpolants. The statement ``assume $\phi$'' is a guard. It aborts
  when~$\phi$ is false. In the transition interpolants, we
  have replaced $v_p$ with $p$ for clarity, but in fact these formulas are over~$V\cup V'$.}
  \label{fig:focisymm}
\end{figure}

\subsection{Modeling with weakest precondition}
A further optimization that we can use in the case of deterministic programs
is that we can express the abstract transition formulas $\Abs{T}$ in
terms of the weakest precondition operator. That is, if~$T$ is deterministic, the abstract transition
formula $\Abs{T}$ is satisfiability equivalent over $V \cup V'$ to:
\[\left({\textstyle \bigwedge_{p\in P}} (v_p \iff p)\right) \wedge \neg \Wp{T}{\False} \wedge \left({\textstyle \bigwedge_{p\in P}} (v'_p \iff \Wp{T}{p})\right)\]
Thus, if we can symbolically compute the weakest precondition operator for 
the operations in our programming language, we can use this formula
in place of $\Abs{T}$ as the abstract transition formula. In this way, the abstract transition formula is localized to just those program variables
that are related in some way to predicates~$P$. In particular, if $\pi$ is an
assignment to a program variable not occurring in $P$, then we will have
$v'_p \iff p$, for every predicate in~$P$.

\subsection{A hybrid approach}
We can combine transition interpolants with other methods of
approximating the transition relation or the image. For example, given a set of propositions~$V$, the \emph{strongest Cartesian postcondition} $\Scp{T}{\phi}$ of a formula~$\phi$ with respect to
a transition formula~$T$ is the strongest cube~$\psi$ over~$V$ such that $\phi \wedge T$ implies~$\psi'$ (a
cube is a conjunction of literals). In computing the image of a state formula~$\phi$
with respect to an operation $\Approx{r}$ of the approximate program, we can strengthen the result
by conjoining it with the strongest Cartesian postcondition with respect to the corresponding abstract operation~$\Abs{r}$.
Thus, the \emph{hybrid image} of~$\phi$ with respect to transition~$\Approx{r}$ is:
\[\Hi{\Approx{r}}{\phi} \doteq \Sp{\Tof{\Approx{r}}}{\phi} \wedge \Scp{\Tof{\Abs{r}}}{\phi}\]
This set is still an over-approximation of the exact abstract image $\Sp{\Tof{\Abs{r}}}{\phi}$,
so it is sound to use the hybrid image in the reachability computation. This may result
in fewer iterations of the refinement loop.


\section{Computing strong interpolants}
\label{sec:strengthen}

In preliminary tests of the method, we found that transition
interpolants derived from proofs by the method of~\cite{CAV03} 
were often unnecessarily weak. 
For example, we might obtain $(p \wedge q) \impliesarrow (p' \wedge q')$ when the stronger $(p
\impliesarrow p') \wedge (q \impliesarrow q')$ could be proved. 
This slowed convergence substantially. 
The experiments presented in this paper use a modified version of the
method of~\cite{CAV03} which incorporates technique of strengthening
the interpolant obtained from a resolution proof . 
It is important in practice to compute strong transition interpolants,
to reduce the number of refinement iterations needed to compute the transiton
relation approximation.

As an example of this, notice that in Figure~\ref{fig:focisymm}, 
the transition interpolant for the second step
is a conjunction of disjunctions:
\[(a[z] = y \impliesarrow (a[z] = y-1)') \wedge ((x = z)' \impliesarrow x= z)\]
However, other valid interpolants are possible. For example, we might have
obtained a weaker version:
\[(x = z)' \impliesarrow (x = z \wedge (a[z] = y  \impliesarrow (a[z] = y-1)'))\]
This formula has been weakened by pulling one disjunction outside of the
conjunction, though it is still sufficient to rule out this particular
program path. The stronger interpolant has the advantage that it
may be more useful in ruling out other program paths in a more
complex program. Unfortunately, either of these interpolants might be
obtained in practice, depending on the exact order of resolution steps
generated by the prover. The order of resolution steps generated by a
SAT solver depends on the order in which implications are propagated
by the Boolean constraint propagation (BCP) procedure, and is quite
arbitrary. Thus, it is useful in practice to try to adjust the proof before
computing an interpolant, in such a way that a stronger interpolant
results.

To understand this process in detail, it is necessary to understand
the process of generating interpolants from resolution proofs, as
described in~\cite{CAV03}. A full treatment of this subject is beyond the scope
of this paper. However, to gain some intuition about the problem, it
is only necessary to know two things about such interpolants. First,
the interpolant for $(A,B)$ is a Boolean circuit whose structure mirrors the
structure of the resolution proof that refutes $A\wedge B$. Second, resolutions on local atoms
(those not occurring in $B$) generate ``or'' gates,
while resolutions on global atoms (those occurring in $B$) generate
``and'' gates. Thus, if we want to generate a strong interpolant
formula, it would be best to move the local resolutions toward the
hypotheses of the proof, and the global resolutions toward the
conclusion. This effectively moves the ``or'' gates toward the inputs
of the interpolant circuit, and the ``and'' gates toward the output,
thus strengthening the interpolant.

\begin{figure}[htbp]
  \centering
  \includegraphics*[angle=270,viewport=0 0 213 426,width=3.5in]{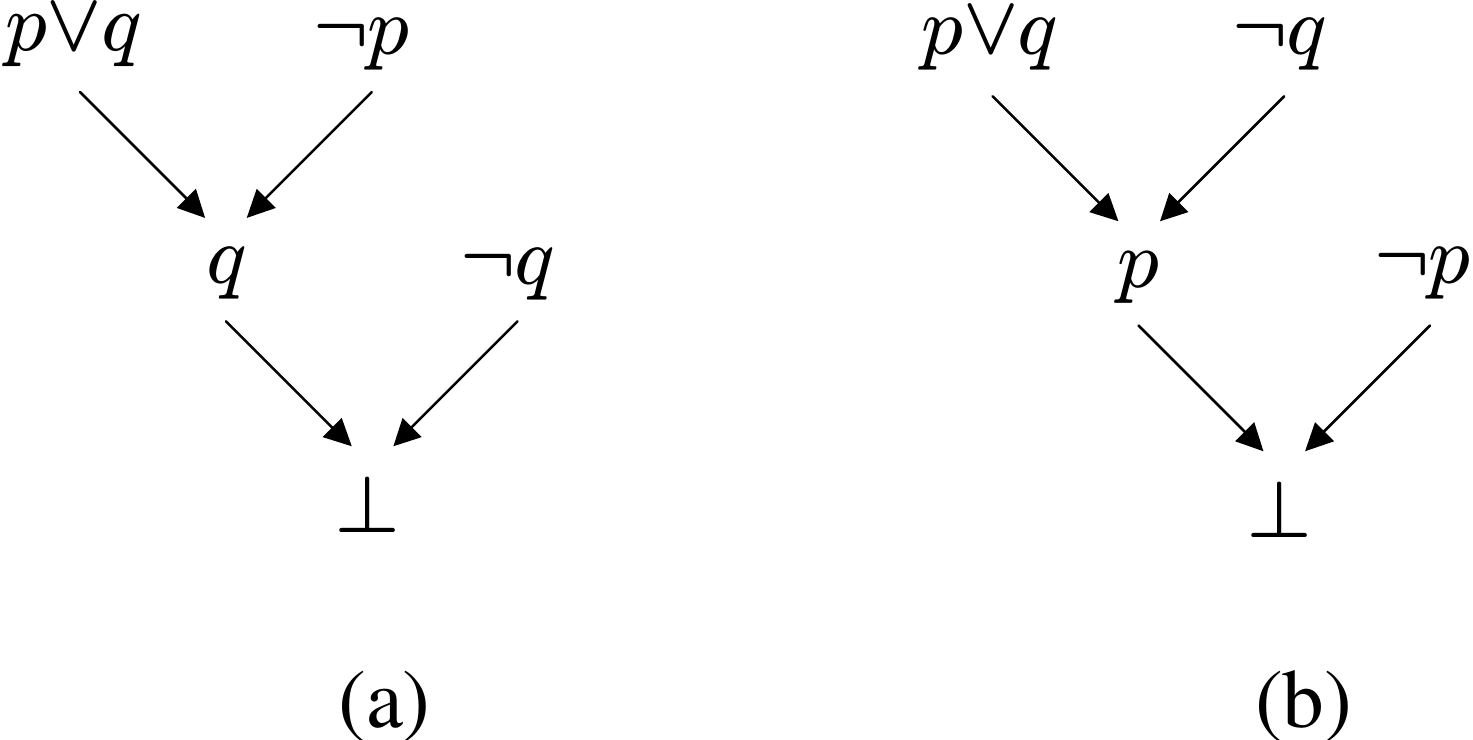}
  \caption{Exchanging the order of two resolution steps}
  \label{fig:resrev1}
\end{figure}

We will think of a refutation proof by resolution as a DAG $(V,E)$, in which the vertices~$V$ are clauses. Each root
of the DAG is a hypothesis of the proof, and the unique leaf is the
empty clause (representing ``false''). Each non-root vertex~$v$ has
exactly two parents, which we will denote $\Antepos{v}$ and
$\Anteneg{v}$, and a \emph{pivot variable} $\Pivot{v}$. The
proof is \emph{valid} when, for every non-root vertex $v$,
$\Antepos{v}$ has the form $\Pivot{v} \vee \Theta_1$ and
$\Anteneg{v}$ has the form $\neg\Pivot{v} \vee \Theta_2$ and $v =
\Theta_1 \vee \Theta_2$ (that is, each derived clause is the result of
resolving its two parents on variable $\Pivot{v}$). Figuratively speaking, each
literal in a hypothesis flows down the DAG until it is annihilated
by resolution with its negation.
As an example,
Figure~\ref{fig:resrev1}a shows a simple resolution refutation
whose hypotheses are $p\vee q$, $\neg p$ and $\neg q$. In this case
we resolve first on $p$, then on $q$. 

\begin{figure}[htbp]
  \centering
  \includegraphics*[angle=270,viewport=0 0 156 417,width=3.5in]{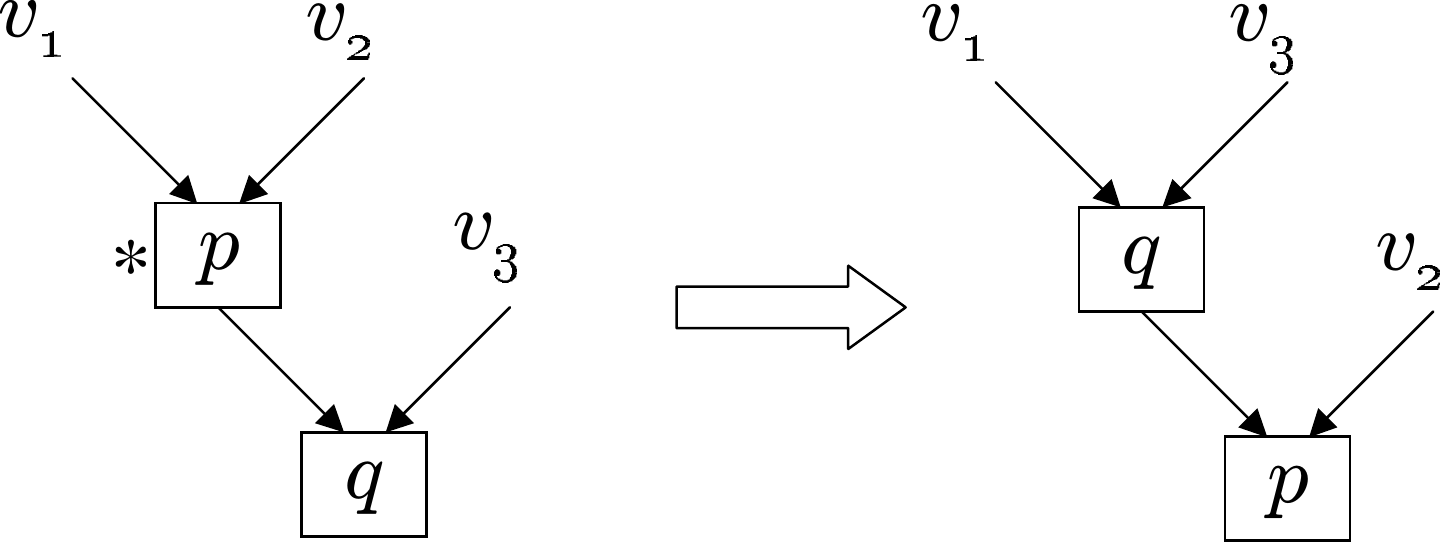}
  \caption{Proof transformation valid when $v_1$ contains $q$, but $v_2$ does not.}
  \label{fig:restrans1}
\end{figure}

The order of two consecutive resolution steps in a proof can always be
reversed, though possibly at the cost of expanding the proof. For
example, Figure~\ref{fig:resrev1}b shows the result of reversing the
order of resolution in our simple example, so that we resolve first
on~$q$, then on~$p$. This is an example of a generally valid graph
transformation on resultion proofs, depicted in Figure~\ref{fig:restrans1}.
In the figure, a box containing a variable~$p$ denotes the result
of resolving its parents on~$p$. The left parent of a vertex~$v$
is $\Antepos{v}$, while the right parent is $\Anteneg{v}$.
This transformation is valid when $q$ occurs in ~$v_1$, but not in~$v_2$.
The reader can easily verify that the right graph is a valid proof when the left
one is. Note that the vertex marked~\asterisk\ on the left hand side
may have other successors in the graph that are not pictured. In this case
we cannot delete this vertex from the proof when applying the transformation.
Thus, applying the transformation may increase the size of the proof graph by
one vertex.

\begin{figure}[htbp]
  \centering
  \includegraphics*[angle=270,viewport=0 0 156 482,width=3.5in]{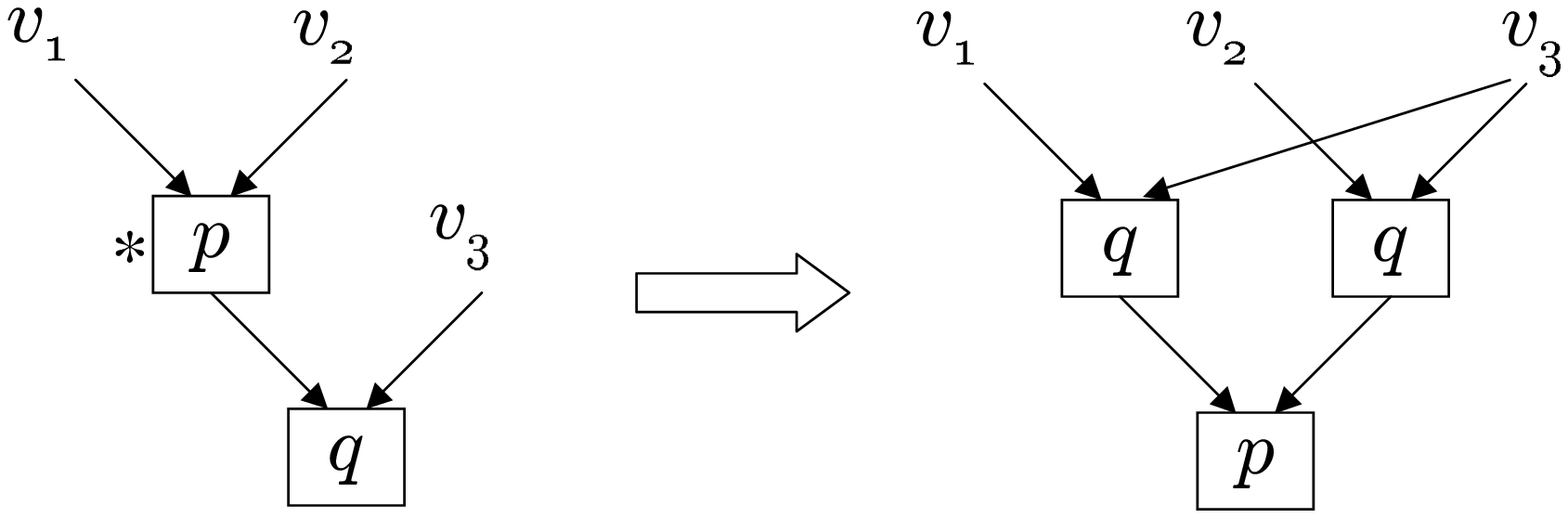}
  \caption{Proof transformation valid when both $v_1$ and $v_2$ contain $q$.}
  \label{fig:restrans2}
\end{figure}

Figure~\ref{fig:restrans2} shows the case when both the clauses~$v_1$ and~$v_2$
contain~$q$. In this case, we obtain two resolutions on~$q$. Again, the original
resolution on~$p$, marked~\asterisk, may or may not be deleted, depending on whether
it has additional successors. Both these transformation are symmetric
with respect to polarity. Thus, we obtain similar transformations by reversing
the antecedents of either resolution step on the left-hand side.

By applying these graph transformations systematically, we can in
principle move all of the resolutions on local atoms to the top of the
proof, and all the resolutions on global atoms to the bottom. This
would result in an interpolant in conjunctive normal form (CNF).
However, it may also result in an exponential expansion of the proof.
Instead, we will take a limited approach that keeps the interpolant
linear in the size of the original resolution proof, but may not yield
an interpolant in CNF.

First, we must first take into account that the proof is a DAG and not a
tree. As noted above, raising resolution on~$q$ above a resolution on~$p$ could
result
in the loss of shared structure, if the latter has more than one successor in the graph.
To prevent this, we never apply the reversal transformations when
the upper resolution (on~$p$) has multiple successors. We will say a vertex is {\em passable}
if it is not a root, and if it has exactly one successor.
We traverse the proof in some topological
order, from antecedents to consequents. Each time we encounter a
resolution step~$v$ on a local atom $q$, we raise this resolution
step by repeatedly exchanging it with one of its parents, until neither
of its parents is passable.
This procedure is shown in pseudocode in Figure~\ref{fig:prooftrans}.

\begin{figure}[htbp]
\begin{tabbing}
mm\=mm\=mm\=mm\=mm\=mm\=mm\=mm\=\kill
function {\sc Resolve}($q,v_1,v_2$)\\
\>   if $q$ does not occur in $v_1$ then return $v_2$\\
\>   else if $\neg q$ does not occur in $v_2$ then return $v_1$\\
\>   else\\
\> \>  let $r$ be the resolvent of $v_1$ and $v_2$ on $q$\\
\> \>  add $r$ to the proof, with $\Antepos{r} = v_1$ and $\Anteneg{r} = v_2$\\
\> \>  {\sc Raise}($r$)\\
\> \>  return $r$\\
\\
procedure {\sc Raise}($v$)\\
\> let $q$ = $\Pivot{v}$\\
\> if $\Antepos{v}$ is passable then\\
\> \> let $v' = \Antepos{v}$ and $p = \Pivot{v'}$\\
\> \> set $\Antepos{v} \leftarrow \mbox{\sc Resolve}(q,\Antepos{v'},\Anteneg v)$\\
\> \> set $\Anteneg{v} \leftarrow \mbox{\sc Resolve}(q,\Anteneg{v'},\Anteneg v)$\\
\> \> remove vertex $v'$ and set $\Pivot{v} \leftarrow p$\\
\> else if $\Anteneg{v}$ is passable then\\
\> \> let $v' = \Anteneg{v}$ and $p = \Pivot{v'}$\\
\> \> set $\Antepos{v} \leftarrow \mbox{\sc Resolve}(q,\Antepos v,\Antepos{v'})$\\
\> \> set $\Anteneg{v} \leftarrow \mbox{\sc Resolve}(q,\Antepos v,\Anteneg{v'})$\\
\> \> remove vertex $v'$ and set $\Pivot{v} \leftarrow p$\\
\\
procedure {\sc TransformProof}\\
\> let $c_1,\ldots,c_n$ be a topological sort of the non-root proof vertices\\
\> for $i = 1\ldots n$ do\\
\> \> if $\Pivot{c_i}$ is local (occurs only in $A$) then {\sc Raise}($c_i$)
\end{tabbing}
\caption{Proof transformation procedure}
\label{fig:prooftrans}
\end{figure}
Note that the procedure {\sc Raise}, if~$q$ occurs in both $\Antepos{v'}$ and
$\Anteneg{v'}$, then we create two new resolutions on~$q$, while removing only one
(this corresponds to the transformation of Figure~\ref{fig:restrans2}).
Thus, the size of the proof increases.
However, the final number of
resolutions on~$q$ is no more than the number of occurrences
of~$q$ in the the original proof.
Thus, the number of resolutions we obtain after raising all the
resolutions on local atoms is linear in the size of the original proof
(if we measure it by the number of literals it contains). As a result
the interpolant we obtain from the rewritten proof is still linear in
size of the original proof (though it may be quadratically larger than
the interpolant derived from the original proof).

Also note that in procedure~{\sc Raise}, it may be possible to raise a
given resolution~$v$ over either $\Antepos{v}$ or $\Anteneg{v}$.  We
have arbitrarily chosen~$\Antepos{v}$ in this case, though it may be,
for example, that the heavier of the two proof branches would be
heuristically the better choice.  This choice occurs rarely, however,
in proofs generated by SAT solvers.  These proofs tend to consist of
long chains of resolutions in which one of the two antecedents are
hypotheses.  These chains are the result of Boolean constraint
propagation. The unusual case in which both antecedents are derived
clauses are typically the result of the SAT solver backtracking out of
a decision. Thus in most cases, only one choice is possible.

There are two reasons why, after apply the transformation procedure, we may still
have global resolutions above local resolutions (and thus ``and''
gates inside ``or'' gates in the interpolant). Most obviously, the
proof may not have been a tree, and thus raising some local resolution may
have been blocked because neither parent was passable. The other reason is that when we raise
resolution on~$q$ above one antecedent, we also raise the proof of the other antecedent.
This may itself contain global resolutions (though as mentioned, in most cases
the other antecedent is a hypothesis)
We might imagine continuing by raising each
resulting resolution on~$q$ above its other antecedent. However, the resulting
loss of structure sharing would cause an exponential expansion in the
proof DAG.  In practice, we have found that the limited transformation
procedure outlined figure~\ref{fig:prooftrans} resutls in an interpolant in CNF
most of the time, producing a substantial improvement in the
performance of interpolation-based refinement over the basic procedure
of~\cite{CAV03}.

\section{Experiments}
\label{sec:results}
We now experimentally compare the method of the previous section with
a method due to Das and Dill~\cite{DBLP:conf/lics/DasD01}. This method refines an approximate transition relation
by analyzing counterexamples from
the approximate system to infer a refinement that rules out each counterexample.
More precisely, a \emph{counterexample} of the
approximate program $(L,\Approx{R})$ is an alternating sequence $\pi =
\sigma_0\Approx{r}_0\sigma_1\cdots\Approx{r}_{k-1}\sigma_k$, where
each $\sigma_i$ is a minterm over $V$, each $\Approx{r}_i$ is an
operation in~$\Approx{R}$, $l(r_0) = l_0$, $l'(r_{k-1}) = l_f$, and for all $0 \leq i < k$, we have
$\Tof{\Approx{r}_i}[\sigma_i,\sigma_{i+1}]$. This induces a set
of \emph{transition minterms}, $t_i = \sigma_i \wedge
\sigma'_{i+1}$, for $0 \leq i < k$.  Note that each $t_i$ is by
definition consistent with $\Tof{\Approx{r}_i}$.
  
To refine the approximate program, we test each $t_i$ for
consistency with the corresponding abstract transition formula
$\Tof{\Abs{r}_i}$. If it is inconsistent, the counterexample is false
(due to over-approximation). Using an incremental decision procedure,
we then greedily remove literals from
$t_i$ that can be removed while retaining inconsistency
with~$\Tof{\Abs{r}_i}$.  The result is a minimal (but not minimum)
cube that is inconsistent with~$\Tof{\Abs{r}_i}$.  The negation of
this cube is implied by $\Tof{\Abs{r}_i}$, so we use it to strengthen
corresponding approximate transition formula~$\Tof{\Approx{r}_i}$.
Since more than one transition minterm may be inconsistent, we may
refine several approximate operations in this way (however if none are
inconsistent, we have found a true counterexample of the abstraction).

Both approximation refinement procedures are embedded as
subroutines of the \Blast\ software model checker.  Whenever the model
checker finds a path from an initial state to a failure state in the
approximate program, it calls the refinement procedure. If refinement
fails because the abstraction does not prove the property, the
procedure of~\cite{POPL04} is used to add predicates to the abstraction.
Since both refinement methods are embedded in the same
model checking procedure and use the same decision procedure, we
can obtain a fairly direct comparison.



Our benchmarks are a set of C programs with assertions embedded to
test properties relating to the contents of arrays.\footnote{
Available at \texttt{http://www-cad.eecs.berkeley.edu/\~{}kenmcmil/cav05data.tar.gz}}
Some of these programs were written expressly as tests. Others were
obtained by adding assertions to a sample
device driver for the Linux operating system from a
textbook~\cite{OReilly}.  Most of the properties are true. None of the properties
can be verified or refuted by \Blast\ without using a refinement
procedure, due to its use of the Cartesian image.

\begin{figure}[t]
  \centering
    \includegraphics[angle=270,width=2.25in]{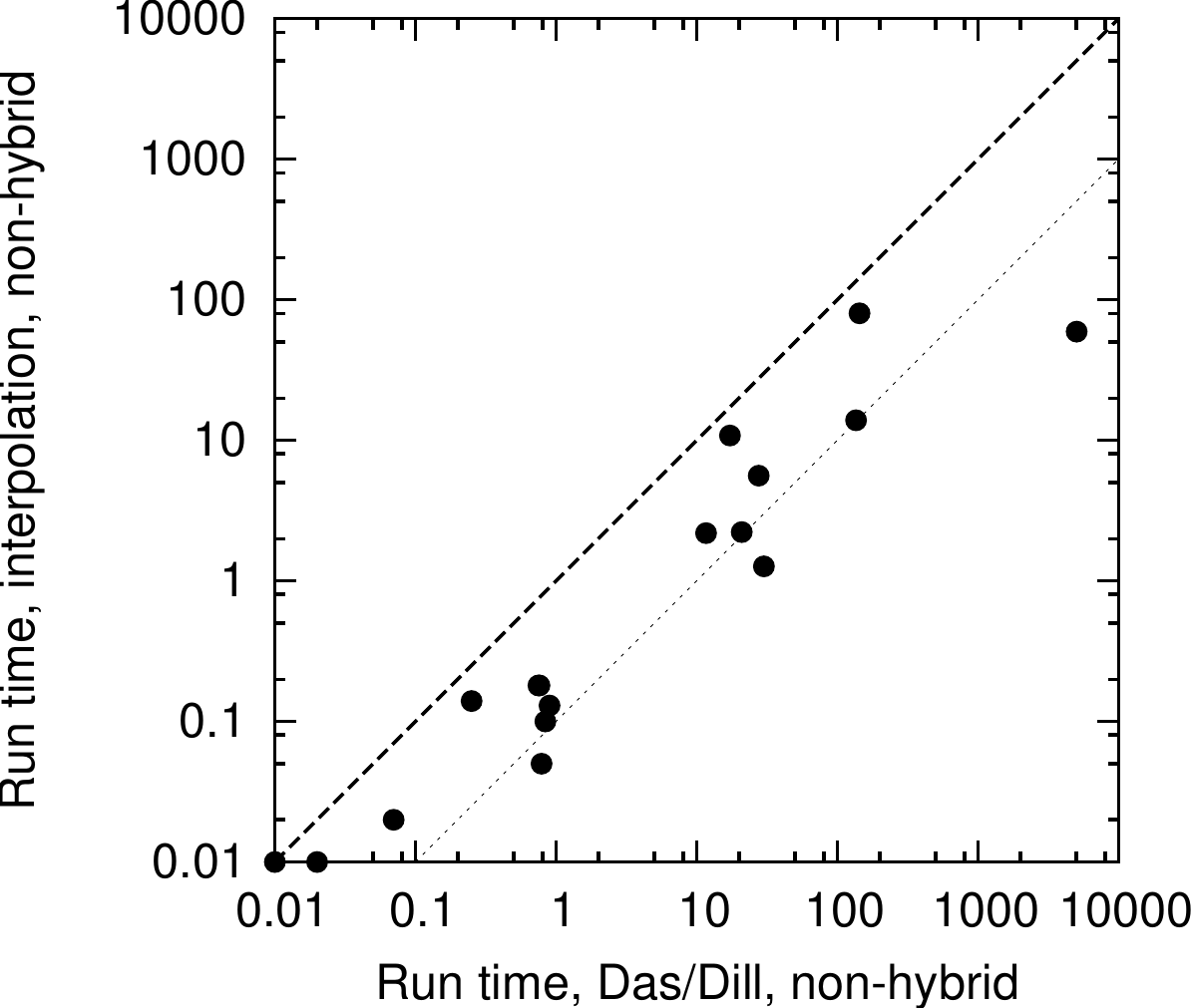} \ \ 
    \includegraphics[angle=270,width=2.25in]{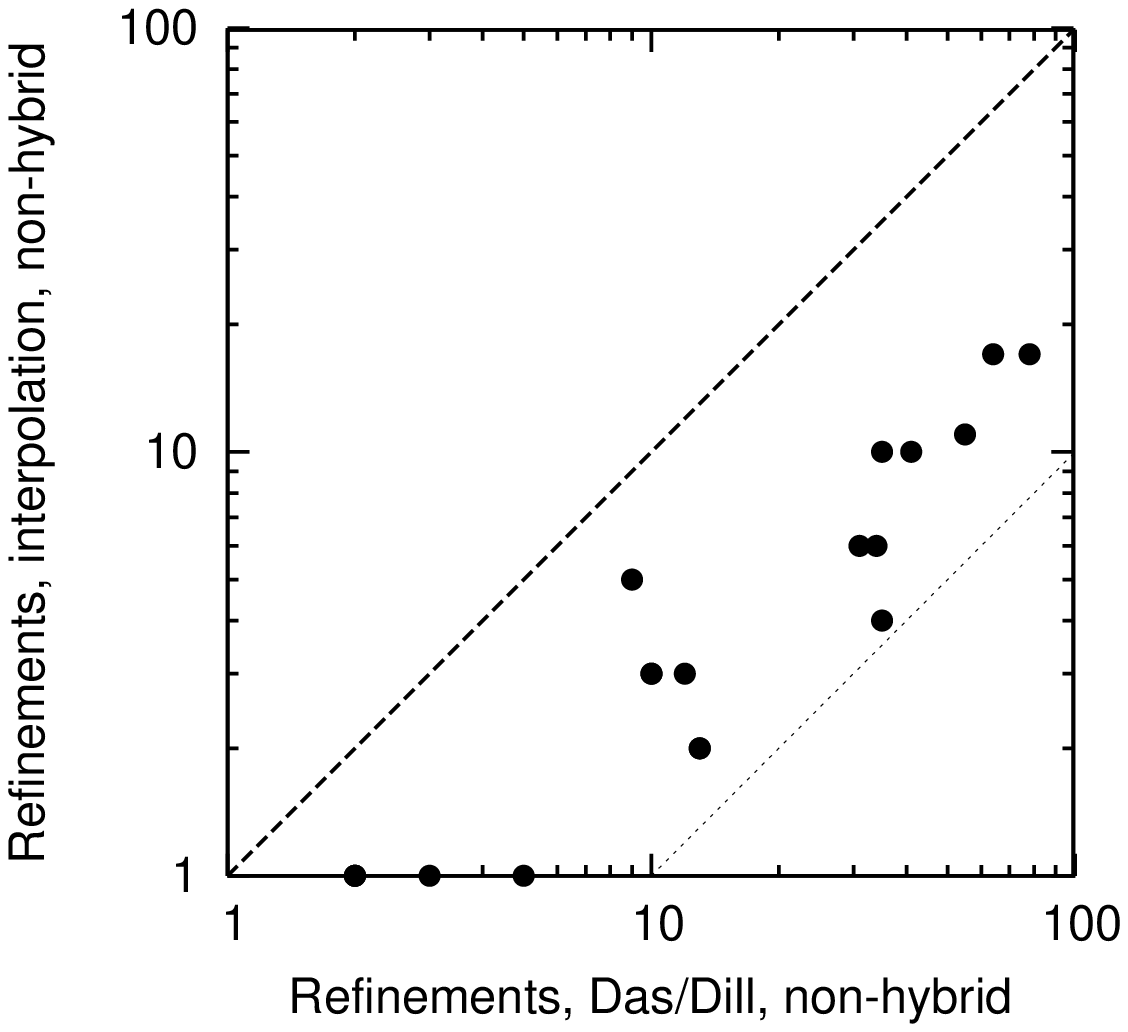}
  \caption{Comparison of the Das/Dill and interpolation-based methods as to run time and number of refinement steps.}
  \label{fig:timesteps}
\end{figure}

\begin{figure}[t]
  \centering
    \includegraphics[angle=270,width=2.25in]{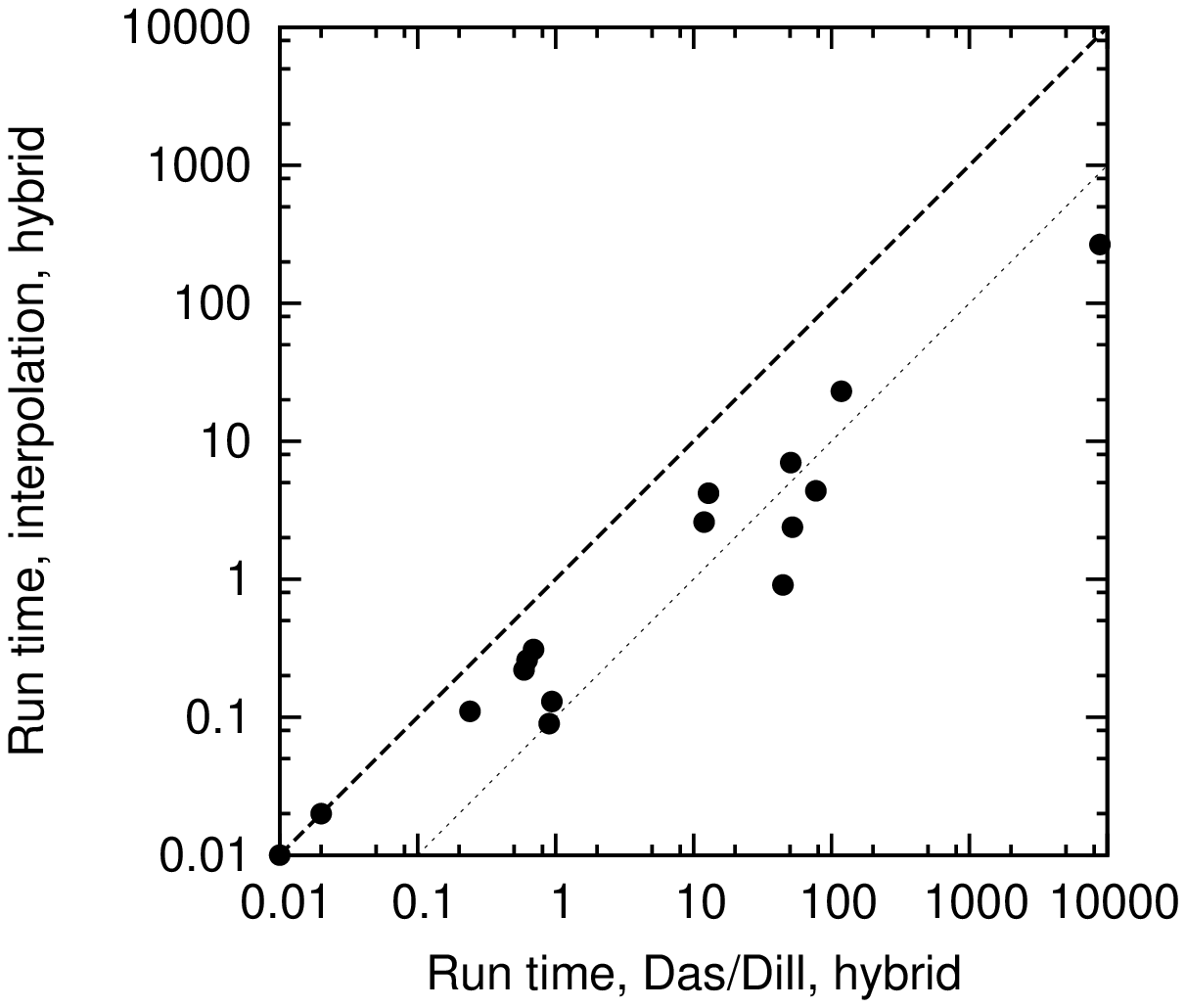} \ \ 
    \includegraphics[angle=270,width=2.1in]{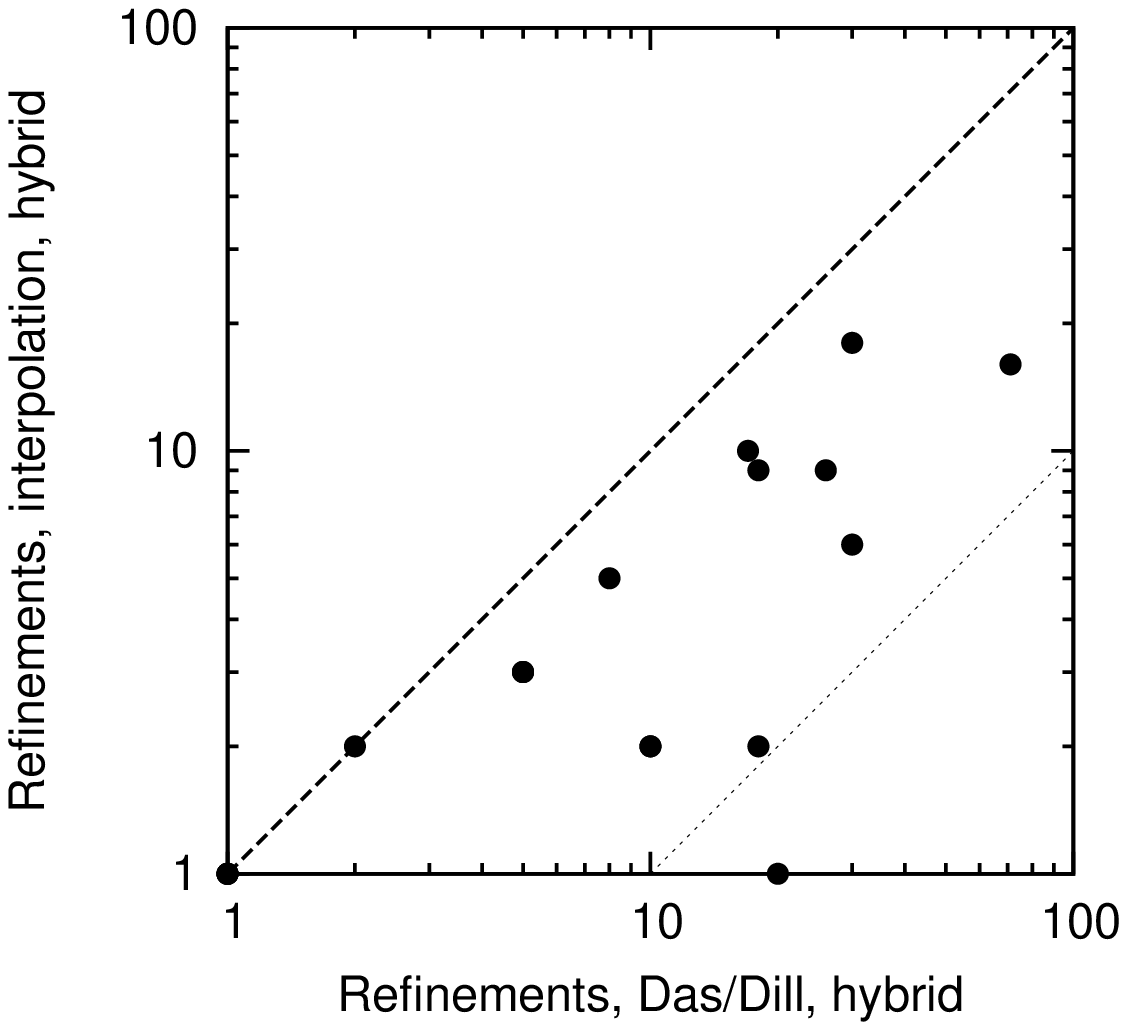}
  \caption{Comparison of the Das/Dill and interpolation-based refinement methods, using the hybrid image.}
  \label{fig:timestepshybrid}
\end{figure}

Figure~\ref{fig:timesteps} shows a comparison in terms of run time (on a 3GHz
Intel Xeon processor) and number of
refinement steps. The latter includes refinement steps that fail, causing
predicates to be added. Run time includes model checking, refinement, and predicate selection.
Each point represents
a single benchmark problem. The~X axis represents the Das/Dill method and the~Y axis the
interpolation-based method.  Points below the heavy diagonal represent
wins for the interpolation method, while points below the light diagonal
represent improvements of an order of magnitude (note in one case a run-time improvement
of two orders of magnitude is obtained).
Figure~\ref{fig:timestepshybrid}
shows the same comparison with the hybrid image computation. Here, the reduction
in number of refinement steps is less pronounced, since less information must be learned by refinement.


The lower number of refinement steps required by interpolation method is
easily explained. The Das/Dill method uses a specific counterexample
and does not consider the property being verified. Thus it can easily
generate refinements not relevant to proving the property. The interpolation
procedure considers only the program path, and generates facts relevant to
proving the property for that path. Thus, it tends to generate more relevant
refinements, and as a result it converges in fewer refinements.

\begin{figure}[t]
  \centering
    \includegraphics[angle=270,width=2.25in]{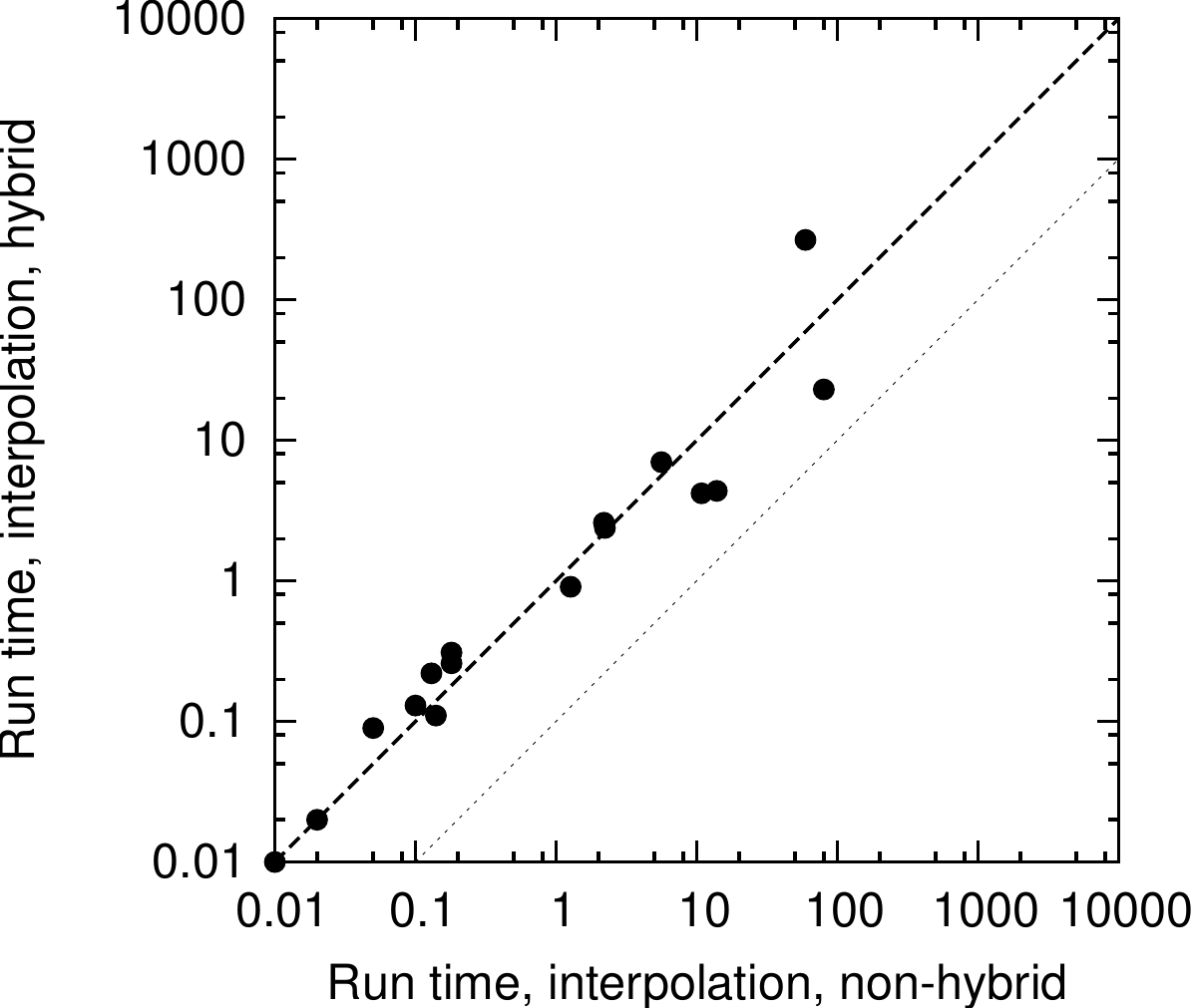} \ \ 
    \includegraphics[angle=270,width=2.1in]{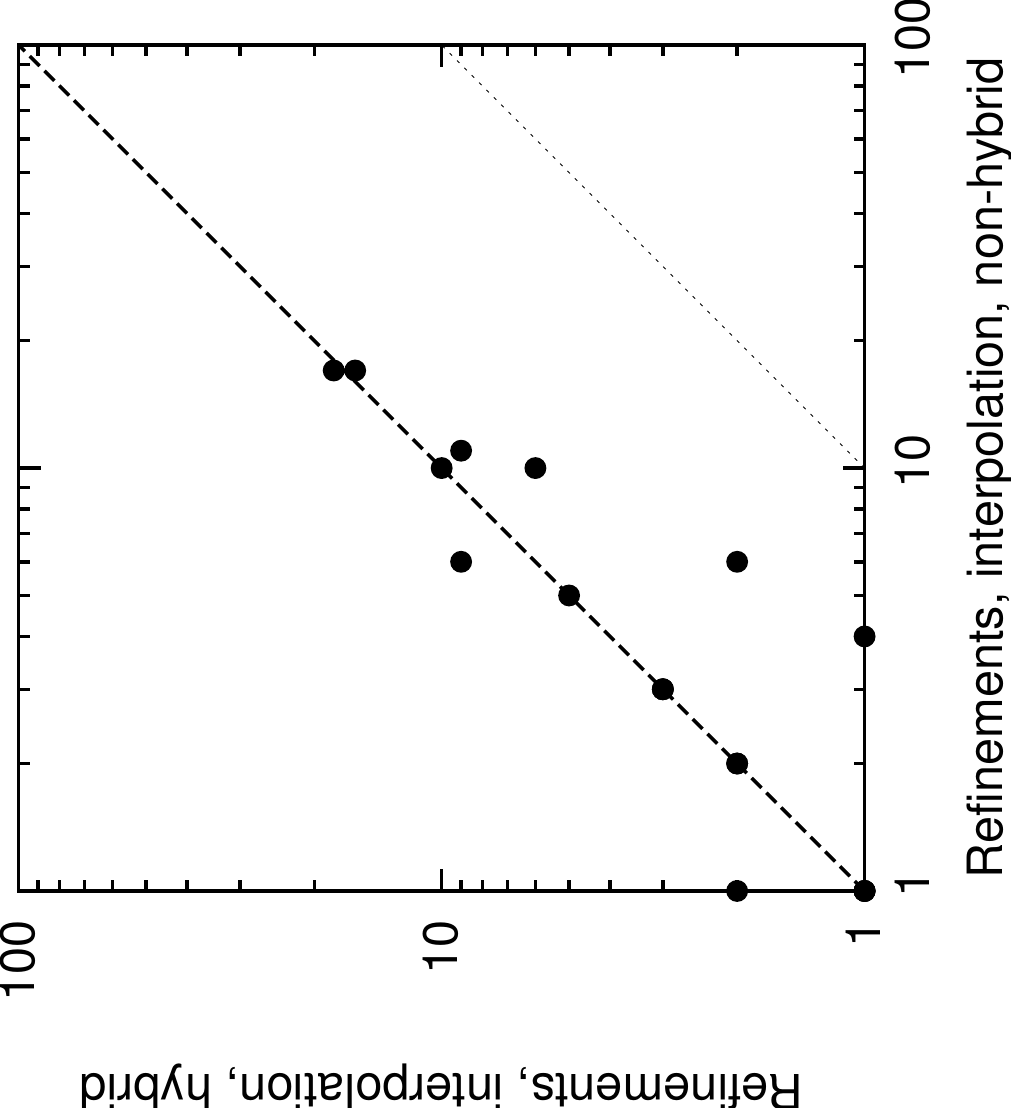}
  \caption{Comparison of the interpolation-based refinement methods, without and with hybrid image.}
  \label{fig:hybrid}
\end{figure}

Figure~\ref{fig:hybrid} compares the performance the interpolation-based method
with and without hybrid image computation. Though the hybrid method can
reduce the number of refinement steps, it sometimes increases the run
time due to the cost of computing the Cartesian image using a decision procedure.


\section{Conclusions}

We have described a method that combines bounded model checking and
interpolation to approximate the transition relation of a system with
respect to a given safety property. The method is extensible to liveness properties of finite-state
systems, in the same manner as the method of~\cite{absref}. When used with predicate
abstraction, the method eliminates the individual variables
and function symbols from the approximate transition formula, leaving it in a propositional
form. Unlike the method of~\cite{SBC2003}, it does this without introducing extraneous
Boolean variables. Thus, we can apply standard symbolic model checking methods
to the approximate system.

For a set of benchmark programs, the method was found to
converge more rapidly than the counterexample-based method of Das and
Dill, primarily due to the prover's ability to focus the proof, and
therefore the refinements, on facts relevant to the property. The
benchmark programs used here are small (the largest being a sample
device driver from a textbook), and the benchmark set contains only~19
problems. Thus we cannot draw broad conclusions about the
applicability of the method.  However, the experiments do show a
potential to speed the convergence of transition relation refinement
for real programs. Our hope is that this will make it easier
to model check data-oriented rather than control-oriented
properties of software.

\bibliographystyle{alpha}
\bibliography{bib}
\vskip-38 pt

\end{document}